\documentclass[superscriptaddress,english,aps,pra,twocolumn,showpacs,superscriptaddress,floatfix]{revtex4}

\usepackage{amsmath}
\usepackage{amssymb}
\usepackage{amsfonts}
\usepackage{bbm}
\usepackage{epsfig}
\usepackage{times}
\usepackage{babel}

\newcommand{\id}{\mathbbm{1}}
\newcommand{\cc}{{\mathbbm{C}}}

\newcommand{\Tr}{\operatorname{Tr}}
\newcommand{\Id}{\operatorname{Id}}
\newcommand{\bra}[1]{\langle {#1}|}
\newcommand{\ket}[1]{|{#1}\rangle}
\newcommand{\Bra}[2]{\,{}_{#2}\langle {#1}|}
\newcommand{\Ket}[2]{|{#1}\rangle_{#2}\,}
\newcommand{\pdag}{{\phantom{\dag}}}
\renewcommand{\vec}[1]{\boldsymbol{#1}}
\newcommand{\mf}[1]{\mathfrak{#1}}
\newcommand{\mc}[1]{\mathcal{#1}}

\newcommand{\vacuum}{\text{\o}}
\newcommand{\F}{\mathcal{F}}
\newcommand{\B}{\mathcal{B}}

\newcommand{\potsdam}{Institute for Physics and Astronomy, Potsdam University, 14476 Potsdam, Germany}
\newcommand{\mexico}{Instituto de F\'{\i}sica, Universidad Nacional Aut\'onoma de M\'exico, M\'exico}

\begin{document}

\title{Contraction of fermionic operator circuits and the simulation of strongly correlated fermions}

\author{Thomas Barthel}
\affiliation{\potsdam}
\author{Carlos Pineda}
\affiliation{\mexico}
\author{Jens Eisert}
\affiliation{\potsdam}

\date{July 20, 2009}
 
\begin{abstract}
A fermionic operator circuit is a product of fermionic operators of usually different and partially overlapping support. Further elements of fermionic operator circuits (FOCs) are partial traces and partial projections.
The presented framework allows for the introduction of fermionic versions of known qudit operator circuits (QUOC), important for the simulation of strongly correlated $d$-dimensional systems: The multiscale entanglement renormalization ans\"atze (MERA), tree tensor networks (TTN), projected entangled pair states (PEPS), or their infinite-size versions (iPEPS etc.).
After the definition of a FOC, we present a method to contract it with the same computation and memory requirements as a corresponding QUOC, for which all fermionic operators are replaced by qudit operators of identical dimension.
A given scheme for contracting the QUOC relates to an analogous scheme for the corresponding fermionic circuit, where additional marginal computational costs arise only from reordering of modes for operators occurring in intermediate stages of the contraction. 
Our result hence generalizes efficient schemes for the simulation of $d$-dimensional spin systems, as MERA, TTN, or PEPS to the fermionic case.
\end{abstract}


\maketitle

\section{Introduction}
Strongly correlated quantum lattice models pose some of the most intriguing physical questions and technical challenges, due to the fact that the number of degrees of freedom increases exponentially with the system size. Classifying the intricacy of calculating ground state energies of such systems has become a vivid branch of complexity theory \cite{Kempe2006-35,Oliveira2008-8,Gottesman2009_05}.
Especially for the analysis of ground state properties in one-dimensional systems, the density-matrix renormalization-group (DMRG) \cite{White1992-11,Schollwoeck2005} provides a numerical approach that is often extraordinarily accurate. It works by variational optimization of a suitable class of states, so-called matrix product states \cite{Accardi1981,Fannes1991,Rommer1997}.
For two- and three-dimensional systems, quantum Monte-Carlo methods (e.g., positive-definite path integral \cite{Suzuki1977-58,Hirsch1982-26} or stochastic series expansion \cite{Sandvik1991-43} representation) are extremely successful for bosonic and unfrustrated spin models, but are bothered by the sign problem \cite{Hirsch1982-26,Takasu1986-75} for some interesting frustrated spin and fermionic models, including the notorious Fermi-Hubbard model
\begin{equation*}
	\hat H= -\sum_{\langle i,j\rangle,\sigma } (\hat f_{i\sigma}^\dag \hat f_{j\sigma}^{\phantom{\dag}}+h.c.)
	+ U\sum_i \hat n_{i\uparrow}\hat n_{i\downarrow} -\mu\sum_{i,\sigma}\hat n_{i\sigma}
\end{equation*}
 which is a candidate for the description of the essential physics of high-temperature superconductivity. Recently, new tools such as the diagrammatic Monte Carlo method have been developed \cite{Prokofev1998-81,VanHoucke2008}, which have a less severe sign problem and have, e.g.,  been demonstrated to give precise results for the repulsive Fermi-Hubbard model in the (correlated) Fermi liquid regime \cite{Kozik2009}.

In a complementary development, generalizations of DMRG ideas to higher dimensions have been put forward. To this purpose, first, one needs to give an ansatz for the many-particle state for which the number of degrees of freedom does only scale polynomial with system size but is (hopefully) still appropriate to describe, e.g., the ground states of the higher-dimensional system. Second, a way of efficiently evaluating interesting local observables or correlators with respect to the ansatz states needs to be identified.
Third, a corresponding algorithm to determine or approximate the ground state within the ansatz class on a classical computer needs to be worked out.
Focusing first on spin (or equivalently qudit) lattices, several suggestions have been put forward, such as {\it tensor product ans\"atze} or {\it projected entangled pair states} (PEPS) \cite{Niggemann1997-104,Nishino2000-575,Martin-Delgado2001-64,Verstraete2004-7,Isacsson2006-74,Verstraete2008-57},
{\it tree tensor networks} (TTN) \cite{Shi2006-74},
or {\it multiscale entanglement renormalization ans\"atze} (MERA) \cite{Vidal-2005-12,Dawson2008-100,Cincio2008-100,Evenbly2009-79,Giovannetti2009-79}.

In this article we address the question of how higher-dimensional \emph{fermionic} systems can be studied via ansatz states. If one maps the system to a spin model by expressing states and operators in the occupation number representation with respect to a fixed ordering of the modes, inevitably long-range ($O(L^{d-1})$, where $L$ is the linear size of the $d$-dimensional lattice) interaction terms occur, rendering simulation unfeasible:
The spin representation of a term $\hat f_j^\dagger\hat f_k$, $j<k$,
under the Jordan-Wigner transformation \cite{Jordan1928} is for instance
\begin{equation*}
	\sigma_j^- \otimes
	\bigotimes_{j<l<k} \sigma_l^z
	 \otimes
	\sigma^+_{k},
\end{equation*}
containing a so-called Jordan-Wigner string.

\begin{figure*}[t]
\centering
\includegraphics[width=0.8\linewidth]{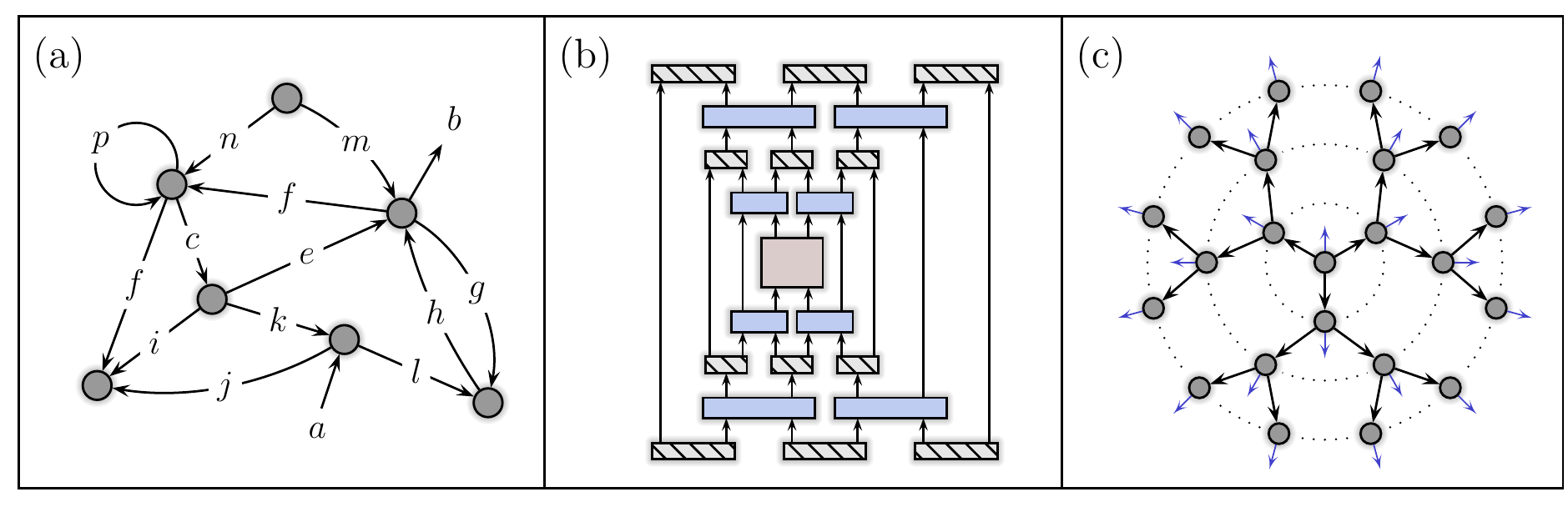}
\caption{(a) The graphical representation of a FOC as a directed graph. The nodes represent fermionic operators. The arcs (directed edges) represent (partial) multiplications, partial traces, and open indices. Each arc is labeled by the set of modes it corresponds to. The operator corresponding to a certain vertex maps from the modes of all incoming arcs to the modes of the outgoing arcs. In the example, the arc ``$e$'' corresponds to a partial multiplication, arc ``$p$'' to a partial trace, and arcs ``$a$'' and ``$b$'' to open incoming and outgoing indices, respectively. The node at the top corresponds to a ket vector from $\F_{m\cup n}$ and the node at the bottom left to a bra vector (element of the dual of $\F_{f\cup i\cup j}$).
As a whole, the circuit is a fermionic operator mapping from $\F_a$ to $\F_b$.
(b) A FOC for the calculation of the expectation value of a local observable (square in the center) with respect to a MERA state with two renormalization steps. The hatched flat rectangles represent isometries which correspond to a coarse graining step in a (realspace) renormalization procedure. The other rectangles represent unitaries that are supposed to reduce entanglement of adjacent blocks before a coarse graining step. The circuit contains only those unitaries and isometries of the MERA that lie inside the so-called causal cone of the observable; all others cancel out.
(c) FOC for a tree tensor network (TTN) state, here for a genuine tree system, the Bethe lattice with coordination number $z=3$. To have the value of a FOC well-defined, one needs to specify an ordering among the operators, assigning to each operator a number $\tau=1,2,\dotsc$. In example (a), we arbitrarily chose  $\tau$ to increase from the bottom to the top. In example (b), a natural ordering, motivated by the picture of subsequent renormalization steps, is also directed from the bottom to the top; as we will explain later, the ordering inside one layer is irrelevant, as the contained isometries are all parity-preserving and operate on disjoint sets of modes. Analogously in example (c), we can choose $\tau$ to increase in radial direction, starting from the central node.}
\label{fig:foc}
\end{figure*}
Accompanied by first numeric results, very recently, fermionic generalizations of MERA states were suggested in Refs.\ \cite{Corboz2009_04,Pineda2009_05} and for PEPS in Ref.~\cite{Kraus2009_04}.
Specifically, in Ref.\ \cite{Pineda2009_05}, also an algorithm for fermionic MERA is given by \emph{dynamical reordering}. It exploits the possibility to change the ordering of the fermionic modes during the algorithm to confine all occurring Jordan-Wigner strings to a sublattice of finite extent, the \emph{causal cone} of, e.g., a local observable in the MERA. Going beyond that result, here, we pose the question whether a given general circuit of fermionic operators (FOC, examples in Fig.~\ref{fig:foc}) can be contracted with the \emph{same} efficiency as a corresponding circuit of qudit operators (QUOC). This is answered in the affirmative for the case where each operator in the FOC is \emph{parity-symmetric} (either fermion number parity preserving or changing): We show constructively that the elementary contraction operations for such a FOC can be executed in an arbitrary sequence and give a detailed account of the algorithm.
As compared to the requirements for the contraction of a certain QUOC with a given contraction scheme, the number of operations and memory requirements for the same contraction scheme, applied to a corresponding FOC, increase only by a marginal amount.

This allows to translate the algorithms already developed for spin systems (for PEPS, e.g., in Refs.\ \cite{Nishino2000-575,Verstraete2004-7,Gu2008-78,Jordan2008-101,Orus2009_05}, for MERA, e.g., in Refs.\ \cite{Dawson2008-100,Rizzi2008-77,Evenbly2009-79,Giovannetti2009-79}) to the fermionic case without loss of computational efficiency.
Giving further details for the case of PEPS, we argue that application of the FOC scheme to fermionic PEPS appears to provide a more efficient algorithm than that presented in Ref.\ \cite{Kraus2009_04} where a mapping to a spin system was employed by choice of a fixed mode ordering.

In Sec.~\ref{sec:FOC} the idea of the FOC is introduced and it is given a proper definition. Rules for the execution of the elementary contraction operations for two or one operators are derived in Sec.~\ref{sec:contractions}, after which the importance of a predefined order among the operators constituting the FOC is pointed out in Sec.~\ref{sec:operatorOrder}. It is also explained how this operator order can be modified with marginal computational cost, allowing to efficiently execute the elementary contractions in an arbitrary sequence. The implications on computational efficiency and locality considerations are summarized in Sec.~\ref{sec:costs}. Sec.~\ref{sec:furtherOp} introduces further useful operations on FOCs that are employed in an efficient contraction algorithm for fermionic (i)PEPS in Sec.~\ref{sec:PEPS}. The article closes with a short discussion.

\section{Fermionic operator circuit}\label{sec:FOC}
\subsection{General structure}
A fermionic operator circuit (FOC) is a product of (not necessarily physical, i.e., in general not particle number parity preserving) fermionic operators $\hat A_i:\F_m\to\F_n$ of in general different support, specified by sets of mode labels $m,n\subset\cal L$.
Further elements of FOCs are partial traces and partial projections.
Each mode label $x\in\mc L$ occurs at most twice, once for an incoming mode of some operator and, the second time, for an outgoing mode of the same or another operator. This means for graphical representations of FOCs as graphs, where each vertex corresponds to one operator $\hat A_i$, that each arc (directed edge) of the graph carries a set of unique mode labels. As explained in Sec.~\ref{sec:FOC-definition} this convention allows for a convenient definition of the FOC such that it has a well-defined value.

Prominent examples of FOCs are fermionic versions of known qudit operator circuits (QUOC), important for the simulation of strongly correlated $d$-dimensional systems:
\emph{multiscale entanglement renormalization ans\"atze} (MERA) \cite{Vidal-2005-12} and \emph{tree tensor networks} (TTN) \cite{Shi2006-74}; Fig.~\ref{fig:foc}. As we show in Sec.~\ref{sec:PEPS} also the fermionic variants of {\it tensor product ans\"atze} or {\it projected entangled pair states} (PEPS) \cite{Niggemann1997-104,Nishino2000-575,Martin-Delgado2001-64,Verstraete2004-7} are covered in the FOC framework; Fig.~\ref{fig:fPEPS_a}.
For a MERA, a possible choice for mode labels are the renormalization step $\tau$ combined with a site label from the corresponding lattice.

For numerical purposes, each fermionic operator $\hat A:\F_{m}\to\F_{n}$ of the circuit is stored in an occupation number representation with respect to certain orderings $\mf{m}$ and $\mf{n}$ of the sets of modes $m,n\subset\mc L$. We consider such orderings as \emph{bijective enumerations} $\mf m:\{1,\dots,|m|\}\to m$ and $\mf n:\{1,\dots,|n|\}\to n$ of the sets, where $|m|$ denotes the number of elements in $m$. We may also treat such enumerations as vectors.
For a chosen ordering $\mf{n}$ of the modes in $n$, we denote the basis states of the Fock space $\F_n$ by
\begin{equation}
	\Ket{\vec{n}}{\mf{n}}= \Ket{n_1,\dots,n_{|n|}}{\mf{n}}:= (\hat f_{\mf{n}_1}^\dag)^{n_1}\dots(\hat f_{\mf{n}_{|n|}}^\dag)^{n_{|n|}} \Ket{\vacuum}{n},
\end{equation}
where $\Ket{\vacuum}{n}$ labels the vacuum state of the Fock space $\F_n$ and $\hat f_i$ are the corresponding anticommuting ladder operators with $\{\hat f_i,\hat f_j^\dag\}=\delta_{ij}$. 
The operator $\hat A$ can hence be stored as the complex $2^{|n|}\times2^{|m|}$ matrix  
\begin{equation}\label{eq:JWrep}
	J_{\mf{n},\mf{m}}(\hat A) = \sum_{\vec{n},\vec{m}} |\vec{n})\Bra{\vec{n}}{\mf{n}} \hat A \Ket{\vec{m}}{\mf{m}}(\vec{m}|.
\end{equation}
This is an occupation number representation or Jordan-Wigner transform \cite{Jordan1928} of the operator $\hat A$.
Of course it is also possible to restrict (for each set of modes) to a reduced basis. The only information about the basis states actually needed is their particle number parity; see Sec.~\ref{sec:modeReordering}.
The states occurring in \eqref{eq:JWrep} are elements of different Hilbert spaces:
$\Ket{\vec{m}}{\mf{m}}\in \F_{m}$,
$\Ket{\vec{n}}{\mf{n}}\in \F_{n}$,
$|\vec{m})\in \B_{|m|}$, \text{ and}
$|\vec{n})\in \B_{|n|}$, where $\B_{|n|}$ denotes the $|n|$-qubit Hilbert space 
\begin{equation}
	\B_{|n|}=(\cc^2)^{\otimes |n|}.	
\end{equation}
A similar approach can be used for anyonic systems \cite{Fradkin1989-63}.

\subsection{Definition of a FOC} \label{sec:FOC-definition}
A fermionic operator circuit is specified by a set of fermionic operators $\{\hat A_i:\F_{m_i}\to\F_{n_i}\}$, where each mode label occurs at most twice, once as an incoming mode of an operator $\hat A_i$ and once as an outgoing mode of an operator $\hat A_j$.
Mode labels which occur two times in this fashion imply a (partial) multiplication, Fig.~\ref{fig:operations}a, or (partial) trace, Fig.~\ref{fig:operations}b, of the corresponding operators with respect to that set of modes. Both operations together define a general contraction of two operators, namely contraction of some outgoing modes of $\hat A$ with some incoming modes of $\hat B$ and, simultaneously, of some incoming modes of $\hat A$ with some outgoing modes of $\hat B$; see Fig.~\ref{fig:operations}c.
Mode labels, which occur only once, correspond to modes that the FOC as a whole maps from or maps to.

To have the value of a FOC well-defined, one needs to specify an ordering of the contained operators $\{\hat A_i\}$. The value of the FOC is then defined by the one resulting from doing the contractions in the order $\hat A_N\circ\dotsc\circ\hat A_2\circ\hat A_1$, where ``$\hat B\circ \hat A$'' denotes the contraction of all common modes of the operators $\hat A$ and $\hat B$; see Fig.~\ref{fig:operations}c. As discussed in Sec.~\ref{sec:operatorOrder}, this operation is associative but in general not commutative, $\hat B\circ\hat A\neq \hat A\circ\hat B$.

\subsection{Remarks on the definition} \label{sec:FOC-remarks}
\begin{figure}[t]
\centering
\includegraphics[width=0.86\linewidth]{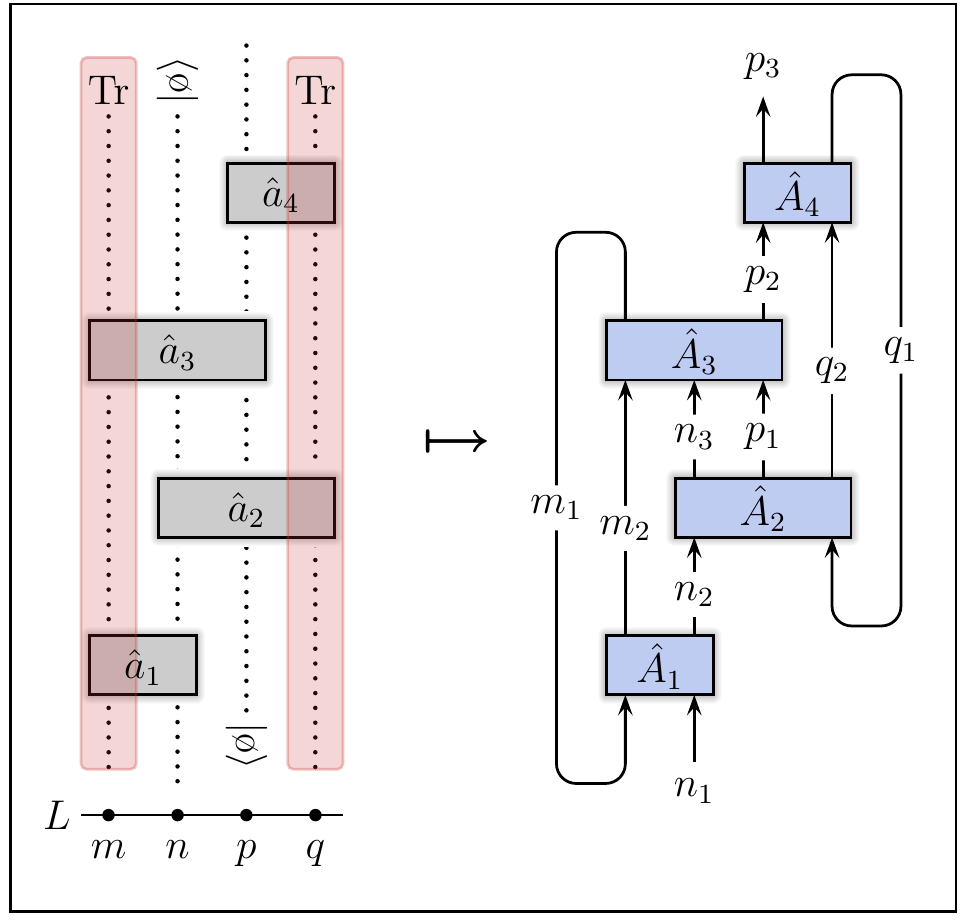}
\caption{(Color online) 
The operator order goes from the bottom to the top. Left: Example for an operator circuit on a lattice $L=m\cup n\cup p\cup q$. It corresponds to the expression $\Tr_{m\cup q}( \Bra{\vacuum}{n} \hat{\mc A}_4 \cdot \dotsc \cdot \hat{\mc A}_1 \Ket{\vacuum}{p})$, cmp.\ Eq.~\eqref{eq:latticeCircuit}, where, e.g., $\hat{\mc A}_1=\hat{a}_1\otimes \Id_{p\cup q}$. For convenience, we require in the definition of fermionic operator circuits, Sec.~\ref{sec:FOC-definition}, that each mode occurs at most twice, once as an incoming mode and once as an outgoing mode of some operators. This can be achieved by a relabeling of the modes, yielding the FOC $\hat A_4\circ\dotsc\circ\hat A_1$ (right). This does not change the matrix elements of the operators and the FOC. One has for example
$\Bra{\vec{m}'\vec{n}'}{\mf{m}^{(2)}\oplus\mf{n}^{(2)}}\hat A_1\Ket{\vec{m}\vec{n}}{\mf{m}^{(1)}\oplus\mf{n}^{(1)}}
=\Bra{\vec{m}'\vec{n}'}{\mf{m}\oplus\mf{n}}\hat a_1\Ket{\vec{m}\vec{n}}{\mf{m}\oplus\mf{n}}$,
where $\mf{m}$, $\mf{n}$, $\mf{m}^{(i)}$, $\mf{n}^{(i)}$ are orderings of the sets of modes $m$, $n$, $m_i$, and $n_i$. Here, with the relabeling, also the partial projections for operators $\hat a_2$ and $\hat a_3$ have been executed.}
\label{fig:latticeCircuit}
\end{figure}
In Sec.~\ref{sec:contractions}, as for the partial contraction operation, we will also give a rule for a partial projection of some modes to basis states (i.e., $\{\hat n_i\}$ eigenstates). This is actually already covered by the contraction operation but perhaps useful to have explicitly, as such projections are frequently used in considerations on operator circuits. 

Note that the operators $\{\hat A_i\}$ are not assumed to be from the so-called algebra of physical operators -- i.e., particle number parity preserving. This is for example useful when calculating correlators of the form $\langle \hat f_i^\dag\hat f_j\rangle$ with respect to MERA or TTN states. In such a calculation, the operators $\hat f_i^\dag$ and $\hat f_j$ become (clearly not parity preserving) elements of a FOC.

However, it will be explained in Sec.~\ref{sec:operatorOrder} that in order to be able to do the contraction of the FOC in an arbitrary sequence (necessary to get optimum numerical efficiency), i.e., to be able to deviate from the order $\hat A_N\circ\dotsc\circ\hat A_2\circ\hat A_1$, it is in general necessary that each $\hat A_i$ is either parity preserving or parity changing.

That mode labels are required to be unique is not a limitation. Consider for example an operator circuit that is defined on a lattice $L$ and does not have that property,
\begin{equation}\label{eq:latticeCircuit}
	\Tr_t( \Bra{\vacuum}{o} \hat{\mc A}_N \cdot \dotsc \cdot \hat{\mc A}_2 \cdot \hat{\mc A}_1 \Ket{\vacuum}{i}).
\end{equation}
Here $t\subset L$ denotes a subset of modes that are traced out, and $i$ and $o\subset L$ denote subsets of modes that are projected out; $t\cap (i\cup o)=\emptyset$. The circuit hence maps from $\F_{L\setminus(t\cup i)}$ to $\F_{L\setminus(t\cup o)}$. Each operator $\hat{\mc A}_i$ acts nontrivially on a subset of the modes: $\hat{\mc A}_i=\hat a_i\otimes \Id_{L\setminus \ell_i}$ with $\hat a_i:\F_{\ell_i}\to\F_{\ell_i}$ where $\ell_i\subset L$.
Now, relabeling of the modes to make the modes unique as depicted in Fig.~\ref{fig:latticeCircuit}, does of course not change the matrix elements of the FOC. It yields a proper FOC $\hat A_N\circ \dotsc\circ\hat A_2\circ\hat A_1$, where each operator $\hat A_i$ has the same matrix elements as the corresponding $\hat a_i$ (partial projections onto the vacuum can be executed in the same step, as in our example, or introduced as separate elements of the FOC). The contraction rules in Sec.~\ref{sec:contractions} are constructed such that this FOC and \eqref{eq:latticeCircuit} have the same matrix elements, i.e., are related by a trivial relabeling of incoming and outgoing modes.

\subsection{Rationale behind calculations and derivations}
\begin{itemize}
	\item The fermionic operators are maps from one Fock space of ``incoming modes'' to another (in general unrelated) Fock space of ``outgoing modes''. In general, they are of different dimension.
	\item Each arc (directed edge) in a graphical representation of a FOC corresponds to a set of unique fermionic modes.
	\item Vacuum states are mode specific. Ladder operators of other unrelated modes commute with the vacuum state for other modes. Take for example $n=\{1,2\}$ and $\mf{n}=(1,2)$, then
\begin{eqnarray}
	\hat f_3^\dag \Ket{n_1n_2}{\mf{n}}
&=&\hat f_3^\dag (\hat f_1^\dag)^{n_1} (\hat f_2^\dag)^{n_2} \Ket{\vacuum}{n} \nonumber\\
&=&(-1)^{n_1+n_2} (\hat f_1^\dag)^{n_1} (\hat f_2^\dag)^{n_2} \Ket{\vacuum}{n}\cdot \hat f_3^\dag \nonumber\\
&=&(-1)^{n_1+n_2} \Ket{n_1n_2}{\mf{n}}\cdot  \hat f_3^\dag
\label{eq:permutationTrick}
\end{eqnarray}
The rationale behind this is that if we have an expression $\Bra{\vacuum}{m\cup n} \hat A_m \hat A_n \Ket{\vacuum}{m\cup n}$ for disjoint sets of modes $m$ and $n$, and where $\hat A_m$ and $\hat A_n$ are polynomials in the ladder operators of the modes in $m$ and $n$, respectively, we have
\begin{equation*}
	\Bra{\vacuum}{n}\Bra{\vacuum}{m} \hat A_m \hat A_n \Ket{\vacuum}{m}\Ket{\vacuum}{n}
=\Bra{\vacuum}{m} \hat A_m \Ket{\vacuum}{m} \Bra{\vacuum}{n} \hat A_n \Ket{\vacuum}{n}
\end{equation*}
\end{itemize}

\subsection{Notation}
We use the Einstein summation convention, i.e., basis state labels that occur twice in an expression presuppose summation over that basis.

Basis states for a certain set $m\subset\cal L$ of $|m|$ fermionic modes and an ordering $\mf{m}$ of those modes will be denoted by $\Ket{\vec{m}}{\mf{m}}= (\hat F^{\vec{m}}_{\mf{m}})^\dag \Ket{\vacuum}{m}$, where $\vec{m}\in \{0,1\}^{|m|}$ and
\begin{equation*}
 \hat F^{\vec{m}}_{\mf{m}}:= (\hat f_{\mf{m}_{|m|}})^{m_{|m|}}\cdot\dotsc\cdot (\hat f_{\mf{m}_2})^{m_2}(\hat f_{\mf{m}_1})^{m_1}.
\end{equation*}
 The number of particles in a basis state $\Ket{\vec{m}}{\mf{m}}$ is denominated by 
\begin{equation}
\bar{m}:= \sum_i m_i.
\end{equation}
The parity of the basis state is $(-1)^{\bar m}$.

Whenever we refer to Fock spaces for unions of sets of modes, as in $\F_{m\cup n}$, it is implied that those sets of modes are disjoint, i.e., $m\cap n=\emptyset$ in that case.

 With $\hat B\cdot_n\hat A$, a partial multiplication is denoted. Only the outgoing modes $n$ of $\hat A$ are contracted with the corresponding same incoming modes $n$ of $\hat B$. Correspondingly $\Tr_r\hat B$ denotes a partial trace, the contraction of incoming modes $r$ with outgoing modes $r$.
By $\hat B\circ\hat A$, we denote a (partial) contraction of all common incoming/outgoing modes of $\hat A$ with corresponding outgoing/incoming modes of $\hat B$.

\section{Contractions} \label{sec:contractions}
In the following, rules are given for all elementary contraction operations needed during the evaluation of a FOC. No non-local Jordan-Wigner transformations occur. The only reordering of modes necessary is for incoming or outgoing modes of single operators, directly before a partial multiplication, trace etc.\ that they are affected by.

\subsection{Reordering of modes} \label{sec:modeReordering}
\begin{figure}[t]
\centering
\includegraphics[width=0.65\linewidth]{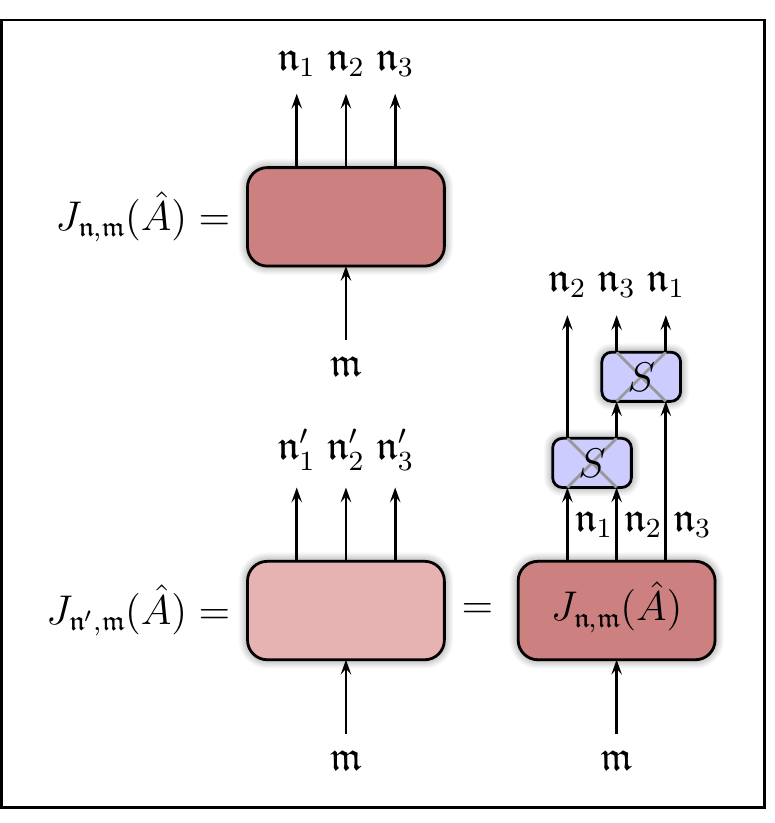}
\caption{(Color online) 
To implement contraction schemes for FOCs on a computer, we represent every operator $\hat A:\F_{m}\to\F_{n}$ in an occupation number representation $J_{\mf{n},\mf{m}}(\hat A)$. The primitive contraction rules, given in Sec.~\ref{sec:contractions}, pose some preconditions on the orderings of modes (to get simple formulae). Hence, before applying those rules, it is in general necessary to change, e.g., from $J_{\mf{n},\mf{m}}(\hat A)$ to a representation $J_{\mf{n}',\mf{m}'}(\hat A)$ with different mode ordering. In the depicted example, the order of the outgoing modes changes from $\mf{n}=(\mf{n}_1,\mf{n}_2,\mf{n}_3)$ to $\mf{n}'=(\mf{n}_2,\mf{n}_3,\mf{n}_1)$. As explained in Sec.~\ref{sec:modeReordering}, this requires application of the swap matrix $S$ [Eq.~\eqref{eq:swapOperator}] -- in this example two times.}
\label{fig:modeReordering}
\end{figure}
\begin{figure*}[t]
\centering
\includegraphics[width=0.9\textwidth]{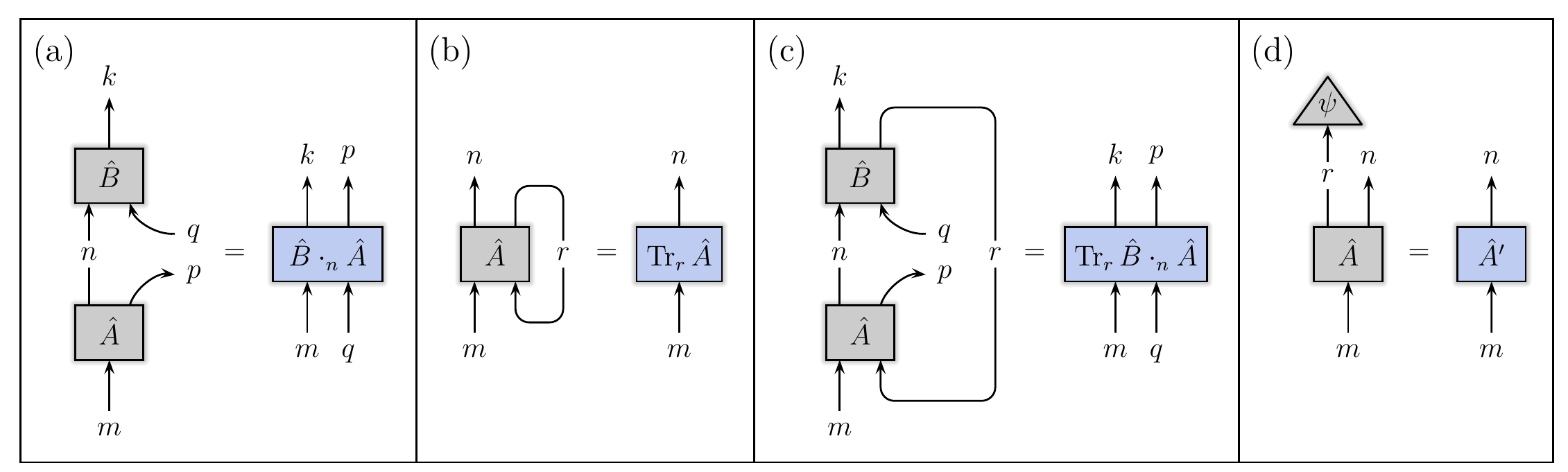}
\caption{(Color online) 
Listing of all (contraction) operations that are needed to evaluate a FOC:
(a) partial multiplication $\hat B\cdot_n\hat A$, (b) partial trace $\Tr_r\hat A$, (c) partial contraction $\hat B\circ\hat A=\Tr_r \hat B\cdot_n\hat A$, and (d) partial projection. The latter two operations are not primitives, but are rather applications of partial multiplication and trace. For numerical purposes, it is however useful to implement them. In all cases, the lower operator is defined to come first in the operator ordering.}
\label{fig:operations}
\end{figure*}
Assume we are given a fermionic operator $\hat A:\F_{m}\to\F_{n}$ in the occupation number representation $J_{\mf{n},\mf{m}}(\hat A)$. The contraction rules to follow, will pose some preconditions on the orderings of modes (to get simple formulae). We need hence to be able to derive from $J_{\mf{n},\mf{m}}(\hat A)$ representations $J_{\mf{n}',\mf{m}'}(\hat A)$ with different mode orders.

All reorderings can be written as sequences of two mode swaps. Let us assume that $\mf{m}'=\mf{m}$ and that orderings $\mf{n}$ and $\mf{n}'$ differ only in modes $\mf{n}_j$ and $\mf{n}_k$ (for $1\leq j<k\leq |n|$), i.e., $\mf{n}_j'=\mf{n}_k$ and $\mf{n}_k'=\mf{n}_j$. Where in the old representation, $|\vec{n})$ corresponds to the state $\Ket{\vec{n}}{\mf{n}}=(\hat f_{\mf{n}_1})^{n_1}\dots (\hat f_{\mf{n}_j})^{n_j} \dots (\hat f_{\mf{n}_k})^{n_k}\dots \Ket{\vacuum}{n}$, it corresponds in the new representation to $\Ket{\vec{n}}{\mf{n}'}=(\hat f_{\mf{n}_1})^{n_1}\dots (\hat f_{\mf{n}_k})^{n_j} \dots (\hat f_{\mf{n}_j})^{n_k}\dots \Ket{\vacuum}{n}$.

To derive the corresponding transformation on the representations of $\hat A$, note that an operator $\hat S_{jk}$ that swaps the modes, i.e., $\hat S_{jk} \hat f_j \hat S_{jk}^\dag= \hat f_k$ and $\hat S_{jk} \hat f_k \hat S_{jk}^\dag= \hat f_j$ is given by \cite{Bravyi2002-1}
\begin{equation}
	\hat S_{jk}=\id-\hat f_j^\dag \hat f^\pdag_j -\hat f_k^\dag \hat f^\pdag_k +\hat f_j^\dag \hat f^\pdag_k + \hat f_k^\dag \hat f^\pdag_j.
\end{equation}
With $\hat S_{jk} \Ket{\vacuum}{n}=\Ket{\vacuum}{n}$, we have hence
\begin{equation}
	J_{\mf{n}',\mf{m}}(\hat A) = J_{\mf{n},\mf{m}}(\hat S_{jk} \hat A)
         = J_{\mf{n},\mf{n}}(\hat S_{jk}) J_{\mf{n},\mf{m}}(\hat A)
\end{equation}
The occupation number representation (Jordan-Wigner transform) of a term $\hat f_j^\dagger
\hat f_k$ is $\sigma^-_j \otimes (\bigotimes_{l=j+1}^{k-1} \sigma_l^z) \otimes \sigma^+_{k}$, where the $\sigma^{\alpha}$ denote the Pauli matrices. The swap operator for two consecutive modes is in the relevant subspace
\begin{eqnarray}
S&:=&J_{(i,i+1),(i,i+1)}(\hat S_{i,i+1})\nonumber\\
&=& \phantom{+}|0,0)(0,0|-|1,1)(1,1|\nonumber\\
&& +|0,1)(1,0| + |1,0)(0,1|.
\label{eq:swapOperator}
\end{eqnarray}
In practice one may choose to execute all mode reorderings by application of corresponding sequences of swap operators for consecutive modes; see Fig.~\ref{fig:modeReordering}.

Swapping of whole sets of modes, e.g., useful when retaining reduced bases, can be done as well. Consider an operator $\hat B:\F_{m}\to\F_{u\cup v\cup x\cup z}$ given in the representation $J_{\mf{n},\mf{m}}(\hat B)$ with $\mf{n}=\mf{u}\oplus\mf{v}\oplus\mf{x}\oplus\mf{z}$ where $\mf{u}$, $\mf{v}$, $\mf{x}$, $\mf{z}$ are orderings for the modes in $u$, $v$, $x$, and $z$. Swapping $\mf{v}$ and $\mf{x}$, is achieved by $(\vec{uxvz}|J_{\mf{n}',\mf{m}}(\hat B)|\vec{m}) = (-1)^{\bar{x}\bar{v}}(\vec{uvxz}|J_{\mf{n},\mf{m}}(\hat B)|\vec{m})$,
where $\mf{n}'=\mf{u}\oplus\mf{x}\oplus\mf{v}\oplus\mf{z}$.

\subsection{Contraction of some outgoing modes of $\hat A$ with the corresponding incoming modes of $\hat B$} \label{sec:multiplyAB}
The partial multiplication of two operators is depicted in Fig.~\ref{fig:operations}a.
Let $\hat A:\F_{m}\to\F_{n\cup p}$ and $\hat B:\F_{n\cup q}\to\F_{k}$, i.e., the operators' outgoing/incoming supports overlap in the modes $n$. Let $\mf{m}$, $\mf{n}$, $\mf{p}$, and $\mf{q}$ be orderings for the modes in $m$, $n$, $p$, and $q$. Assuming we have the two operators in representations $A=J_{\mf{a},\mf{m}}(\hat A)$ and $B=J_{\mf{k},\mf{b}}(\hat B)$ with $\mf{a}=\mf{n}\oplus\mf{p}$ and $\mf{b}=\mf{n}\oplus\mf{q}$, the resulting operator $\hat C:= \hat B\cdot_{n}\hat A:\F_{m\cup q}\to\F_{k\cup p}$ with orderings $\mf{c}_1=\mf{k}\oplus\mf{p}$, $\mf{c}_2=\mf{m}\oplus\mf{q}$ is
\begin{eqnarray}
	\hat C&=&\hat B\cdot_{n}\hat A \nonumber\\
 &=& \Ket{\vec{k}}{\mf{k}}(\vec{k}|B|\vec{n}'\vec{q})\Bra{\vec{n}'\vec{q}}{\mf{b}} \cdot \Ket{\vec{n}\vec{p}}{\mf{a}}(\vec{n}\vec{p}|A|\vec{m})\Bra{\vec{m}}{\mf{m}} \nonumber\\
 &=& (-1)^{\bar{p}\bar{q}+(\bar{p}+\bar{q})(\bar{n}+\bar{n}')}
  \Ket{\vec{k}}{\mf{k}}(\vec{k}|B|\vec{n}'\vec{q}) \nonumber\\
&& \times \Bra{\vacuum}{q}\Bra{\vacuum}{n} 
 (F^{\vec{p}}_\mf{p})^\dag F^{\vec{n}'}_\mf{n} (F^{\vec{n}}_\mf{n})^\dag F^{\vec{q}}_\mf{q}
  \Ket{\vacuum}{n}\Ket{\vacuum}{p} \nonumber\\
&&\times (\vec{n}\vec{p}|A|\vec{m})\Bra{\vec{m}}{\mf{m}} \nonumber\\
 &=&(-1)^{\bar{p}\bar{q}} \cdot \Ket{\vec{k}\vec{p}}{\mf{c}_1}(\vec{k}|B|\vec{n}\vec{q}) (\vec{n}\vec{p}|A|\vec{m}) \Bra{\vec{m}\vec{q}}{\mf{c}_2} \nonumber\\
 &=:& \Ket{\vec{k}\vec{p}}{\mf{c}_1}(\vec{k}\vec{p}|C|\vec{m}\vec{q})\Bra{\vec{m}\vec{q}}{\mf{c}_2},
\label{eq:partialMultiply}
\end{eqnarray}
where $C$ is the representation $C=J_{\mf{c}_1,\mf{c}_2}(\hat C)$. In short, the transformation rule for the occupation number representations reads
\begin{equation}
	(\vec{k}\vec{p}|C|\vec{m}\vec{q}) = (-1)^{\bar{p}\bar{q}}(\vec{k}|B|\vec{n}\vec{q}) (\vec{n}\vec{p}|A|\vec{m}).
\end{equation}
In appendix~\ref{sec:multiplyAB_alternative}, an alternative derivation of this rule is given, where the support of operators $\hat A$ and $\hat B$ is extended prior to the contraction such that there is no need for applying the commutation prescription \eqref{eq:permutationTrick}. The result is the same.

\subsection{Partial trace of an operator}
The partial trace of an operator is depicted in Fig.~\ref{fig:operations}b.
Let $\hat A:\F_{m\cup r}\to\F_{n\cup r}$, i.e., the operator's outgoing and incoming supports overlap in the modes $r$. Such operators can always be decomposed in the form
\begin{equation}
	\hat A=\hat A_+ + \hat A_-,
\end{equation}
where $\hat A_+$ is the particle number parity preserving and $\hat A_-$ the parity changing component, i.e.,
\begin{equation}
	(-1)^{\hat N_n+\hat N_r}\hat A_\pm=  \pm \hat A_\pm(-1)^{\hat N_m+\hat N_r}
\end{equation}
with $\hat N_r:= \sum_{i\in r} \hat f^\dag_i\hat f_i$.

The correct expression for the partial trace follows from its defining property that $\Tr(\hat A\hat B)=\Tr(\Tr_r(\hat A)\hat B)$ for all operators $\hat B$ that have no support on modes $r$. Hence, let us
consider such an operator $\hat B:\F_{n\cup r}\to\F_{m\cup r}$ with no support on $r$, i.e., $\hat f_{i}\hat B_\pm=\pm \hat B_\pm\hat f_{i}$ $\forall_{i\in r}$.
Let $\mf{m}$, $\mf{n}$, $\mf{r}$ be orderings for the modes in $m$, $n$, and $r$. Further let $\mf{a}=\mf{m}\oplus\mf{r}$ and $\mf{b}=\mf{n}\oplus\mf{r}$.
The operator's matrix elements obey
\begin{eqnarray}
	&&\Bra{\vec{m}\vec{r}'}{\mf{a}} \hat B\Ket{\vec{n}\vec{r}}{\mf{b}} \nonumber\\
&&=
 \Bra{\vacuum}{a} \hat F^{\vec{r}'}_\mf{r}\hat F^{\vec{m}}_\mf{m}\hat B (\hat F^{\vec{n}}_\mf{n})^\dag(\hat F^{\vec{r}}_\mf{r})^\dag
 \Ket{\vacuum}{b} \nonumber\\
 &&=(-1)^{\bar{r}'\bar{m}+\bar{r}\bar{n}}
 \Bra{\vacuum}{m}\Bra{\vacuum}{r} \hat F^{\vec{m}}_\mf{m}\hat F^{\vec{r}'}_\mf{r}\hat B (\hat F^{\vec{r}}_\mf{r})^\dag(\hat F^{\vec{n}}_\mf{n})^\dag
 \Ket{\vacuum}{r} \Ket{\vacuum}{n} \nonumber\\
&&=\delta_{\vec{r}\vec{r}'} (-1)^{\bar{r}(\bar{m}+\bar{n})} \Bra{\vec{m}}{\mf{m}} \hat B_+ +(-1)^{\bar{r}}\hat B_-\Ket{\vec{n}}{\mf{n}} \nonumber\\
&&=\delta_{\vec{r}\vec{r}'} \Bra{\vec{m}}{\mf{m}} \hat B\Ket{\vec{n}}{\mf{n}}.
\label{eq:matrixElement}
\end{eqnarray}
Requiring that
\begin{equation*}
	\Tr(\hat A\hat B) 
         = \Bra{\vec{n}\vec{r}}{\mf{b}} \hat A \Ket{\vec{m}\vec{r}}{\mf{a}} \Bra{\vec{m}}{\mf{m}} \hat B\Ket{\vec{n}}{\mf{n}}
	 =\Tr((\Tr_r \hat A)\hat B), 
\end{equation*}
is true for all operators $\hat B$ with the properties stated above, leads to the conclusion that the partial trace for the modes $r$ is simply given by the expression
\begin{equation}\label{eq:partialTrace}
	\Tr_r \hat A = \sum_{\vec{r}} 
 \Ket{\vec{n}}{\mf{n}}\Bra{\vec{n}\vec{r}}{\mf{b}} \hat A \Ket{\vec{m}\vec{r}}{\mf{a}}\Bra{\vec{m}}{\mf{m}}.
\end{equation}
Hence, assuming we have the operator in the representation $J_{\mf{b},\mf{a}}(\hat A)$, the resulting operator $\Tr_r \hat A:\F_{m}\to\F_{n}$ is in the occupation number representation
\begin{equation}\label{eq:partialTraceJW}
	(\vec{n}|J_{\mf{n},\mf{m}}(\Tr_r \hat A )|\vec{m}) =  (\vec{n}\vec{r}|J_{\mf{b},\mf{a}}(\hat A)|\vec{m}\vec{r}).
\end{equation}

Please note that we have chosen the orderings of the modes such that, in Eq.~\eqref{eq:matrixElement}, two sign factors compensate -- that of a mode reordering with one from commuting $\hat F^{\vec{r}}_\mf{r}$ and the operator $\hat B$.
A sign factor $(-1)^{\bar{r}(\bar{m}+\bar{n})}$ would occur in the expressions for the partial trace, had we swapped the order of $\mf{m}$ ($\mf{n}$) and $\mf{r}$ in the ordering of the incoming (outgoing) modes, i.e., $\mf{a}=\mf{r}\oplus\mf{m}$ ($\mf{b}=\mf{r}\oplus\mf{n}$) instead of our choice here. For such a case, the preparative mode reordering would take account of the sign factor and then, having realized the preconditions of it, one would apply rule \eqref{eq:partialTraceJW}.

\subsection{Contraction of some outgoing modes of $\hat A$ with the corresponding incoming modes of $\hat B$ and vice versa}
Combining partial multiplication \eqref{eq:partialMultiply} with partial trace \eqref{eq:partialTrace} we obtain a general partial contraction, namely, that of some outgoing modes $n$ of operator $\hat A$ with the corresponding incoming modes of $\hat B$ and, simultaneously, contraction of some outgoing modes $r$ of $\hat B$ with the corresponding incoming modes of $\hat A$.
This corresponds to the partial contraction depicted in Fig.~\ref{fig:operations}c.

Let $\hat A:\F_{m\cup r}\to\F_{n\cup p}$ and $\hat B:\F_{n\cup q}\to\F_{k\cup r}$, i.e., the operators outgoing/incoming supports overlap in the modes $n$ and $r$.
Let $\mf{m}$, $\mf{n}$, $\mf{r}$, $\mf{p}$, $\mf{q}$, $\mf{k}$ be orderings for the modes in $m$, $n$, $r$, $p$, $q$, and $k$. Assuming we have the two operators in representations $A=J_{\mf{n}\oplus\mf{p},\mf{m}\oplus\mf{r}}(\hat A)$ and $B=J_{\mf{k}\oplus\mf{r},\mf{n}\oplus\mf{q}}(\hat B)$, with $\mf{a}=\mf{k}\oplus\mf{p}$ and $\mf{b}=\mf{m}\oplus\mf{q}$,
the resulting operator $\hat C:\F_{m\cup q}\to\F_{k\cup p}$ is
\begin{eqnarray*}
	\hat C&=&\Tr_r\hat B\cdot_{n}\hat A \nonumber\\
 &=&(-1)^{\bar{p}\bar{q}+\bar{r}(\bar{p}+\bar{q})}\cdot \Ket{\vec{k}\vec{p}}{\mf{a}}(\vec{k}\vec{r}|B|\vec{n}\vec{q}) (\vec{n}\vec{p}|A|\vec{m}\vec{r}) \Bra{\vec{m}\vec{q}}{\mf{b}},
\end{eqnarray*}
i.e.,
\begin{multline}
(\vec{k}\vec{p}|J_{{\mf{a}},{\mf{b}}}(\hat C)|\vec{m}\vec{q})\\
=(-1)^{\bar{p}\bar{q}+\bar{r}(\bar{p}+\bar{q})}\cdot (\vec{k}\vec{r}|B|\vec{n}\vec{q}) (\vec{n}\vec{p}|A|\vec{m}\vec{r}).
\label{eq:partialContract}
\end{multline}
In the following, $\hat B\circ\hat A$ denotes a (partial) contraction of all common incoming/outgoing modes of $\hat A$ with corresponding outgoing/incoming modes of $\hat B$ according to Eq.~\eqref{eq:partialContract}.

\subsection{Partial projection}\label{sec:partialProjection}
The partial projection for an operator is depicted in Fig.~\ref{fig:operations}d.
Let $\hat A:\F_{m}\to\F_{r\cup n}$.
Let $\mf{r}$, $\mf{m}$, $\mf{n}$ be orderings for the modes in $r$, $m$, and $n$. Further let $\mf{a}=\mf{r}\oplus\mf{n}$.
After projection of modes $r$ onto a basis state ($\{\hat n_i\}_{i\in r}$ eigenstate) $\Ket{\vec{r}'}{\mf{r}}= (\hat F^{\vec{r}'}_\mf{r})^\dag \Ket{\vacuum}{r}$, the resulting operator $\hat A':\F_{m}\to\F_{n}$ is
\begin{eqnarray}
	\hat A'&=&\Bra{\vec{r}'}{\mf{r}} \cdot \Ket{\vec{r}\vec{n}}{\mf{a}}(\vec{r}\vec{n}|J_{\mf{a},\mf{m}}(\hat A)|\vec{m})\Bra{\vec{m}}{\mf{m}} \nonumber\\
&=& \Ket{\vec{n}}{\mf{n}}(\vec{r}'\vec{n}|J_{\mf{a},\mf{m}}(\hat A)|\vec{m})\Bra{\vec{m}}{\mf{m}},
\end{eqnarray}
i.e.,
\begin{equation}
	(\vec{n}|J_{\mf{n},\mf{m}}(\hat A')|\vec{m}) = (\vec{r}'\vec{n}|J_{\mf{a},\mf{m}}(\hat A)|\vec{m}).
\label{eq:partialProjection}
\end{equation}
A sign factor $(-1)^{\bar{r}'\bar{n}}$ would occur, if we would swap the order of modes $\mf{r}$ and $\mf{n}$ in the order $\mf{a}$ of the outgoing modes.

\begin{figure}[t]
\centering
\includegraphics[width=0.55\linewidth]{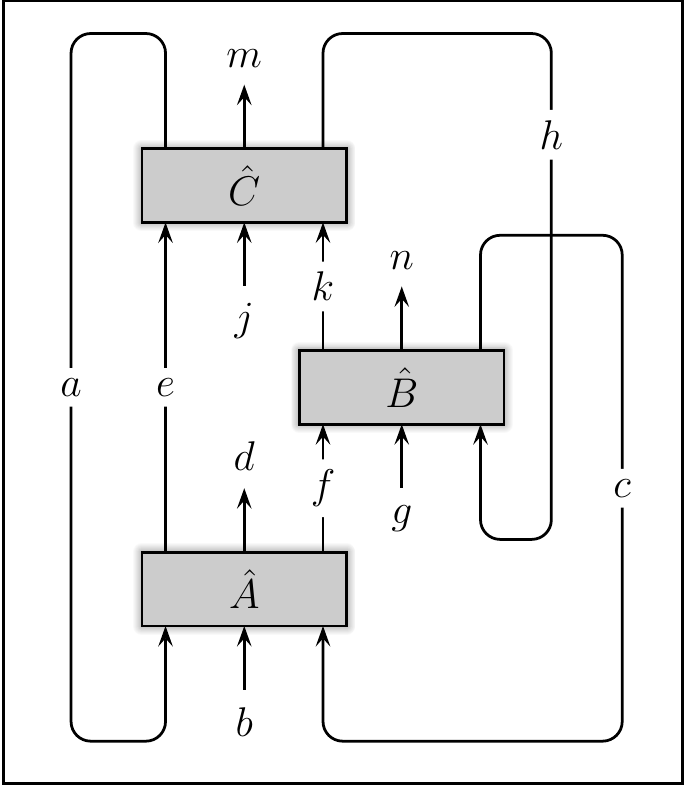}
\caption{The most general FOC with three operators. To verify that the contraction of operators as given by rule \eqref{eq:partialContract} is associative, one needs to compare the results of $\hat C\circ(\hat B\circ\hat A)$ and $(\hat C\circ\hat B)\circ\hat A$. Both do indeed agree.}
\label{fig:associativity}
\end{figure}
\section{Operator order and contraction sequence}\label{sec:operatorOrder}
\begin{figure*}[t]
\centering
\includegraphics[width=\textwidth]{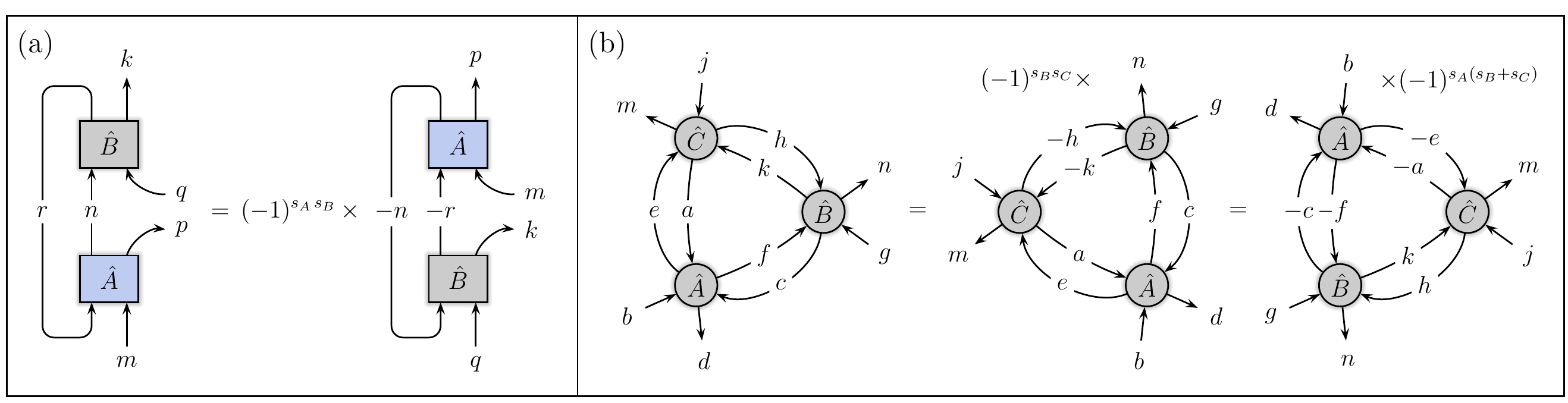}
\caption{(Color online) 
To allow for arbitrary contraction sequences, one needs to be able to change the operator ordering. In the diagrams, the operator order is defined to increase from the bottom to the top. If each operator is parity-symmetric (either preserves or changes the fermion number parity; $s=0$ or $s=1$), swapping of operators can be done and the resulting sign factors taken account of efficiently. (a) The generic rule \eqref{eq:operatorOrderSwap} for swapping two operators that are neighbors in the ordering. (b) Identities for the most generic FOC with three operators, the same as in Fig.~\ref{fig:associativity}, depicted in a slightly different fashion. A minus sign at the contraction arc for a mode set $n$ indicates that a sign factor $(-1)^{\bar n}$ is to be inserted in the contraction formula (see text).}
\label{fig:operatorReordering}
\end{figure*}
In Sec.~\ref{sec:FOC-definition}, the value of the FOC was defined as the value resulting from executing the contractions of the constituting operators $\hat A_i$ with respect to a certain operator order, $\hat A_N\circ\dotsc\circ\hat A_2\circ\hat A_1$. This definition is only sufficient if the contraction \eqref{eq:partialContract} of operators, as depicted in Fig.~\ref{fig:operations}c, is indeed associative. For the most general FOC of three operators $\hat A:\F_{a\cup b\cup c}\to\F_{d\cup e\cup f}$, $\hat B:\F_{f\cup g\cup h}\to\F_{c\cup k\cup n}$, and $\hat C:\F_{e\cup j\cup k}\to\F_{a\cup h\cup m}$ one finds indeed (see Fig.~\ref{fig:associativity})
\begin{equation}
	\hat C\circ(\hat B\circ\hat A) = (\hat C\circ\hat B)\circ\hat A,
\end{equation}
confirming the consistency of the contraction rule \eqref{eq:partialContract}.

Numerically it may be more efficient to execute for example first the contraction between $\hat A_1$ and $\hat A_3$ and contract the result with $\hat A_2$ afterwards. To be able to choose an arbitrary sequence for the contractions as is possible for the corresponding QUOCs, we need to be able to change the ordering of the operators without changing the value of the FOC.
In the elementary contractions, the ordering of the affected operators matters, i.e., for two operators $\hat A:\F_{m\cup r}\to\F_{n\cup p}$ and $\hat B:\F_{n\cup q}\to\F_{k\cup r}$, we have in general $\Tr_r \hat B\cdot_n \hat A \neq \Tr_n \hat A\cdot_r \hat B$. However, if each of the two operators is either parity preserving ($s=0$) or parity changing ($s=1$), we find the simple relation
\begin{equation} \label{eq:operatorOrderSwap}
	\Tr_r \hat B\cdot_n \hat A = (-1)^{s_A s_B}\Tr_n[ (\hat P_n \cdot_n \hat A \cdot_r \hat P_r )\cdot_r \hat B],
\end{equation}
where $\hat P_n: \F_n\to \F_n $ with $\Bra{\vec{n}'}{\mf{n}}\hat P_n\Ket{\vec{n}}{\mf{n}}=\delta_{\vec{n}\vec{n}'}(-1)^{\bar n}$. In the more compact notation this reads $\hat B\circ\hat A = (-1)^{s_A s_B} \hat P_n \circ \hat A \circ \hat P_r \circ \hat B$. In an implementation, instead of inserting the $\hat P_n$ in this fashion as operators or applying them directly to $\hat A$ or $\hat B$, more efficiently, one may introduce a binary counter (with initial state 0) for each contraction arc -- in this case, for the contraction with respect to modes $n$. Whenever a factor $\hat P_n$ arises when swapping the order of operators that have both support on $n$, the state of the binary counter is inverted. Once, the contraction with respect to modes $n$ is executed, one inserts the factor $(-1)^{\bar{n}}$ in the corresponding expression, if the state of the counter is 1. In the graphical representation of a FOC, we denote the state 1 of the counter by a minus sign at the corresponding contraction arc as exemplified in Fig.~\ref{fig:operatorReordering}. The numerical overhead for keeping track of those signs is marginal.

In the following, operators $\hat A:\F_m\to\F_n$ that are either fermion number parity preserving or changing,
\begin{equation} \label{eq:paritySymmetric}
(-1)^{\hat N_n}\hat A=\pm\hat A(-1)^{\hat N_m},
\end{equation}
are called \emph{parity-symmetric}. Also FOCs that contain only parity-symmetric operators are called parity-symmetric.

Using the above result, it is possible to do the operator contractions of a parity-symmetric FOC in an arbitrary sequence. One starts with the predefined operator order. To execute the contraction of two (arbitrary) operators of the FOC:
\begin{itemize}
  \item Apply rule \eqref{eq:operatorOrderSwap} to bring the two operators into direct neighborhood in the operator order, keeping track of the resulting sign factors for the contraction arcs and of the global sign,
  \item apply mode swapping operators as described in Sec.~\ref{sec:modeReordering}, to bring the occupation number representations of the two operators into accord with the precondition of the general contraction rule \eqref{eq:partialContract}, and
  \item replace the two operators by their contraction according to the rule \eqref{eq:partialContract}.
\end{itemize}

Consequently, the contraction of a FOC can be done efficiently -- with the same sequence of partial contractions as for a corresponding qudit operator circuit. No non-local Jordan-Wigner transformations occur. Marginal computational overheads result from keeping track of certain sign factors when doing contractions in a sequence that deviates from the ordering of the circuit's operators and reordering of modes for incoming or outgoing modes of single operators, directly before a partial multiplication, trace etc.\ that they are affected by.

The operator order is part of the definition of a FOC. For the example of the fermionic MERA it can be chosen to agree with the physical interpretation as consecutive renormalization steps; i.e., the operator order is increasing with the renormalization number. As all unitaries (isometries) of a particular renormalization stage commute, the ordering among those can be chosen arbitrarily. In Sec.~\ref{sec:PEPS} a useful operator ordering for fermionic PEPS is presented.

\section{Computational costs and locality} \label{sec:costs}
Given a contraction sequence for a qudit operator circuit (QUOC), the same sequence can be used for a corresponding parity-symmetric FOC (for which all qudit operators are replaced by parity-symmetric fermionic operators of identical dimension). There is hence no memory or computational overhead \emph{per se}. For the elementary contraction operations stated in Sec.~\ref{sec:contractions},  a certain ordering of the modes was being assumed, prior to the operation.
If one uses the contraction operations as stated there, one gets a marginal overhead from the corresponding preparative mode reorderings; Sec.~\ref{sec:modeReordering}. The number of numerical operations needed for a reordering is proportional to the size of the operator matrix: every reordering can be achieved by a sequence of swaps of consecutive modes. The product of appropriate swaps yields a reordering operator that is sparse with exactly one entry $\pm 1$ in each row and column. To apply such an operator to either side of $\hat A:\F_{m}\to\F_{n}$, requires only $\chi_m\chi_n$ operations, where $\chi_m$ and $\chi_n$ are the dimensions of the (possibly reduced) incoming and outgoing Hilbert spaces.
Every contraction of the operator, except for partial traces or projections, would however already require a larger number of numerical operations. The computational overhead is hence marginal. There is no overhead in memory requirements.

Further, all considerations about locality, hence, carry over directly from those of the known QUOCs (for instance the qudit MERA) to the corresponding FOC (e.g., the fermionic MERA).
In the calculation of local expectation values w.r.t.\ a MERA, only operators inside a causal cone of the observable enter the actual calculation (all others cancel). That Jordan-Wigner strings outside the causal cone can be avoided for the fermionic MERA has already been shown by an alternative approach in Ref.\ \cite{Pineda2009_05}, see also Ref.\ \cite{Corboz2009_04}.

\section{Further operations on FOCs} \label{sec:furtherOp}
\begin{figure}[t]
\centering
\includegraphics[width=0.92\linewidth]{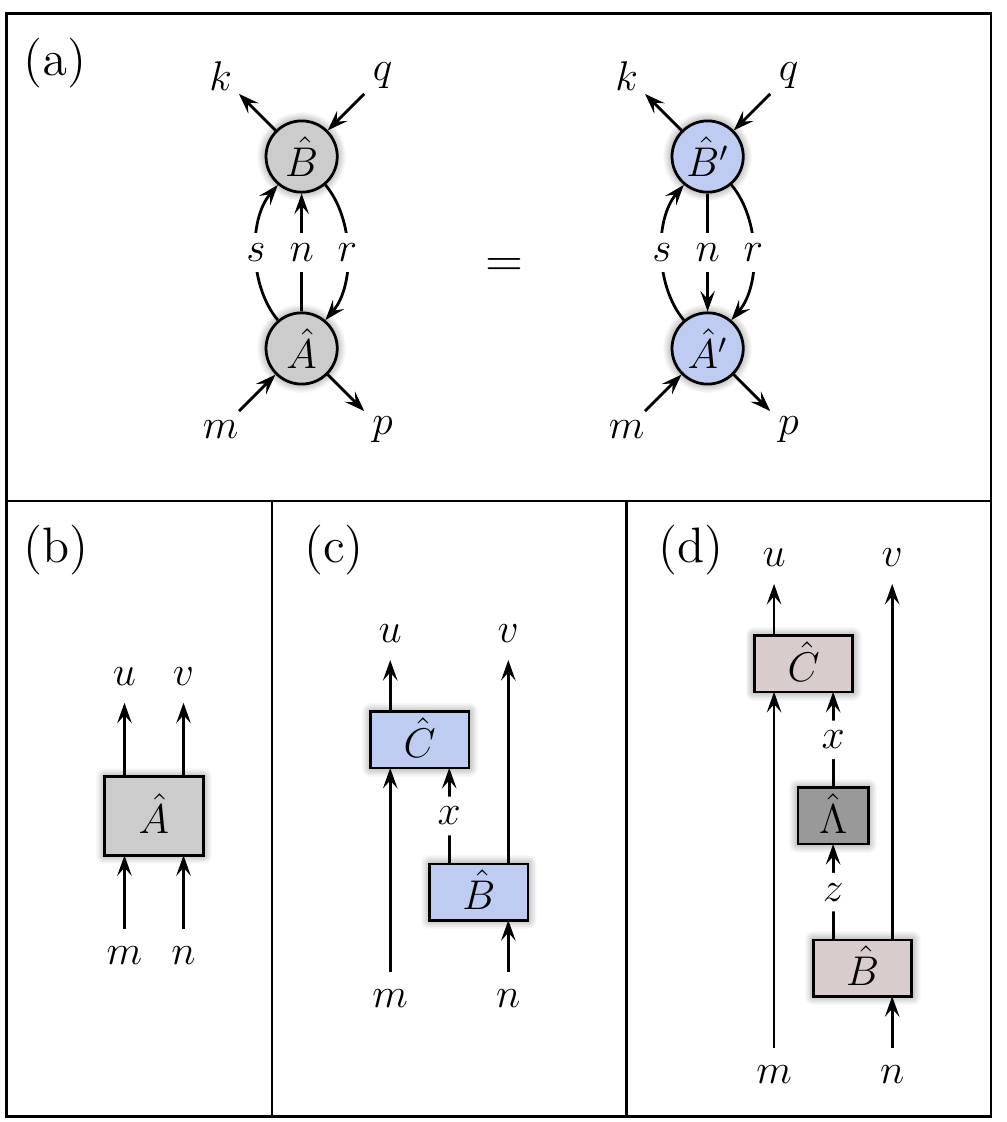}
\caption{(Color online) 
In all subplots, the operator order is defined to increase from the bottom to the top.
(a) It is possible to reverse contraction arcs. The resulting operators can be expressed in terms of matrix elements of the original operators; see Sec.~\ref{sec:reversing}. Reversing the arc for modes $n$ yields the sign factor $(-1)^{\bar n(\bar p+\bar q)}$.
(b--d) It is possible to decompose operators ($\hat A$) by singular value decomposition, resulting in circuits of the form (c) or (d). This also allows for the reduction of retained Hilbert space dimensions: Contract operators $\hat B$ and $\hat C$ to obtain an operator $\hat A$, apply the singular value decomposition to it and truncate (some of the smallest) singular values, to obtain an approximation of $\hat C\circ\hat B$.
Reversing contraction arcs and truncation of Hilbert spaces via singular value decomposition are for example employed in the contraction algorithm for fermionic PEPS in Sec.~\ref{sec:PEPS}.}
\label{fig:reverse-n-svd}
\end{figure}

\subsection{Hermitian conjugation}\label{sec:herm}
The Hermitian conjugate of a FOC is simply given by
\begin{equation}
 (\hat A_N\circ\dotsc\circ\hat A_1)^\dag=\hat A_1^\dag\circ\dotsc\circ\hat A_N^\dag .
\end{equation}
The operator order is reversed and one has to take the Hermitian conjugate of each fermionic operator in the circuit. In the representation as a directed graph, all arcs are reversed.
The Hermitian conjugate is for example of interest when calculating expectation values with respect to a (pure) FOC state. Fig.~\ref{fig:fPEPS_a}a shows it for the example of a fermionic PEPS.

\subsection{Reversing contraction arcs}\label{sec:reversing}
For algorithms operating on FOCs, as for example the one for fermionic PEPS presented in Sec.~\ref{sec:PEPS}, it is sometimes useful to reverse contraction arcs, i.e., to change outgoing modes of one operator to incoming modes and vice versa at the operators it is contracted with; see Fig.~\ref{fig:reverse-n-svd}a. Let $\hat A:\F_{m\cup r}\to\F_{n\cup s\cup p}$ and $\hat B:\F_{n\cup s\cup q}\to\F_{k\cup r}$, i.e., the operators outgoing/incoming supports overlap in the modes $n$, $r$ and $s$.
Let $\mf{m}$, $\mf{n}$, $\mf{r}$, $\mf{s}$, $\mf{p}$, $\mf{q}$, $\mf{k}$ be orderings for the modes in  $m$, $n$, $r$, $s$, $p$, $q$, and $k$.
For reversing the arc corresponding to modes $n$, i.e., changing the modes $n$ to be incoming (outgoing) at operator $\hat A$ ($\hat B$), the relations between $\hat A$ and $\hat B$ and the resulting operators (as depicted in Fig.~\ref{fig:reverse-n-svd}a) are
\begin{gather*}
	\Bra{\vec{kr}}{\mf{k}\oplus\mf{r}}\hat B\Ket{\vec{nsq}}{\mf{n}\oplus\mf{s}\oplus\mf{q}}
= (-1)^{\bar n\bar q}
    \Bra{\vec{knr}}{\mf{k}\oplus\mf{n}\oplus\mf{r}}\hat B'\Ket{\vec{sq}}{\mf{s}\oplus\mf{q}},\\
	\Bra{\vec{nsp}}{\mf{n}\oplus\mf{s}\oplus\mf{p}}\hat A\Ket{\vec{mr}}{\mf{m}\oplus\mf{r}}
= (-1)^{\bar n\bar p} 
    \Bra{\vec{sp}}{\mf{s}\oplus\mf{p}}\hat A'\Ket{\vec{mnr}}{\mf{m}\oplus\mf{n}\oplus\mf{r}},
\end{gather*}
such that $\hat B\circ\hat A = \hat B'\circ\hat A'$.

\subsection{Singular value decomposition and truncation}\label{sec:svd}
It is possible to decompose an operator $\hat A:\F_{m\cup n}\to\F_{u\cup v}$ by singular value decomposition with respect to arbitrary splittings of the incoming and outgoing modes. The resulting circuits can be chosen to be of the form $\hat C\circ\hat B$ or $\hat C\circ\hat \Lambda\circ\hat B$, where $\hat \Lambda:\F_z\to\F_x$ ($|x|=|z|$) is a diagonal operator encoding the singular values; see Figs.~\ref{fig:reverse-n-svd}b--\ref{fig:reverse-n-svd}d.
This also allows for truncation of modes (or the reduction of Hilbert space dimensions): Contract two operators $\hat C:\F_{m\cup x}\to\F_{u}$ and $\hat B:\F_{n}\to\F_{x\cup v}$, as in Fig.~\ref{fig:reverse-n-svd}c to obtain an operator $\hat A$, apply the singular value decomposition to it and truncate (some of the smallest) singular values, to obtain an approximation of $\hat C\circ\hat B$ where the dimension of the retained Hilbert space for the modes in $x$ has been reduced.

Let $\mf{m}$, $\mf{n}$, $\mf{u}$, $\mf{v}$, $\mf{x}$, and $\mf{z}$ be orderings of modes in $m$, $n$,  $u$, $v$, $x$, and $z$.
The contraction of the FOC $\hat C\circ \hat B$, as depicted in Fig.~\ref{fig:reverse-n-svd}c yields
\begin{multline}
	\hat C\circ \hat B = (-1)^{\bar m\bar v}\Ket{\vec{uv}}{\mf{u}\oplus\mf{v}}\Bra{\vec{u}}{\mf{u}}\hat C\Ket{\vec{xn}}{\mf{x}\oplus\mf{n}}\\
\times \Bra{\vec{xv}}{\mf{x}\oplus\mf{v}}\hat B\Ket{\vec{n}}{\mf{n}}\Bra{\vec{nm}}{\mf{n}\oplus\mf{m}}.
\end{multline}
With the occupation number representation $A:=J_{\mf{u}\oplus\mf{v},\mf{n}\oplus\mf{m}}(\hat A)$ of $\hat A$, we can hence decompose the operator by applying the singular value decomposition to the matrix $\tilde A$ defined by
\begin{gather}
	(\vec{uv}|\tilde A|\vec{nm}) := (-1)^{\bar m\bar v}(\vec{uv}|A|\vec{nm}),\\
\tilde A = U \Lambda V,
\end{gather}
where $U$ and $V$ are unitary and $\Lambda$ is the diagonal matrix of singular values. The operators of the resulting circuit $\hat C\circ \hat B$ can then be chosen as ($0<\alpha<1$)
\begin{equation}
	J_{\mf{u},\mf{x}\oplus\mf{m}}(\hat C)=U\Lambda^{\alpha},\quad 
	J_{\mf{x}\oplus\mf{v},\mf{n}}(\hat B)=\Lambda^{1-\alpha}V.
\end{equation}
When the singular values are to be separated into a third operator $\hat \Lambda$ as depicted in Fig.~\ref{fig:reverse-n-svd}d, the operators of the resulting circuit $\hat C\circ\hat \Lambda \circ\hat B$ are given by
\begin{equation}
	J_{\mf{u},\mf{x}\oplus\mf{m}}(\hat C)=U,\quad
        J_{\mf{x},\mf{z}}(\hat \Lambda) = \Lambda,\quad
	J_{\mf{z}\oplus\mf{v},\mf{n}}(\hat B)=V.
\end{equation}

Reduction of Hilbert space dimensions (\emph{truncation}) via singular value decomposition is for example employed in the algorithm for evaluating expectation values with respect to fermionic PEPS in an approximative fashion; see Sec.~\ref{sec:PEPS}.

\section{Fermionic PEPS}\label{sec:PEPS}
\begin{figure*}[p]
\centering
\includegraphics[width=\textwidth]{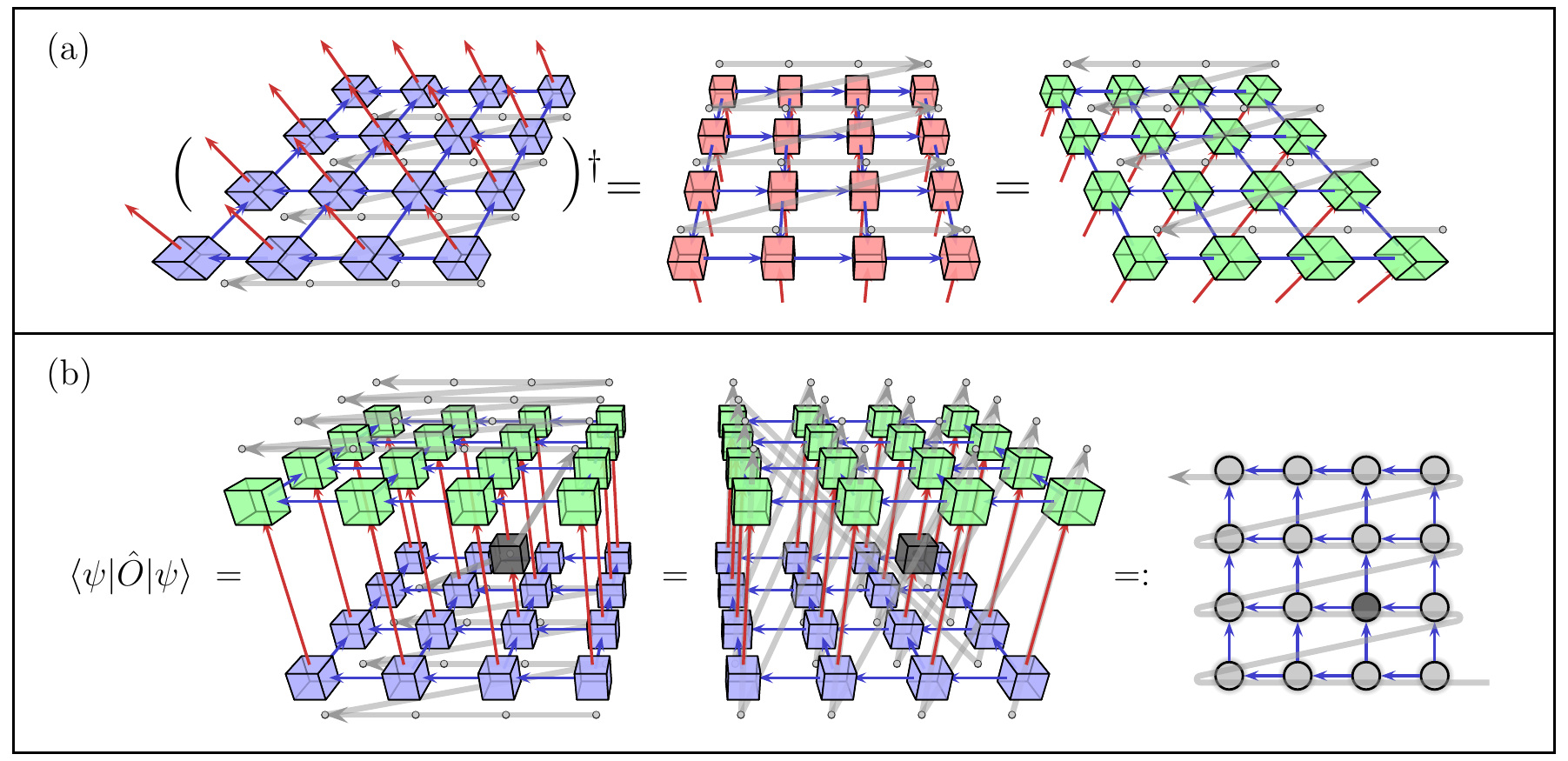}
\caption{(Color online) 
(a) A fermionic PEPS can be constructed as a FOC, where fermionic operators are assigned to each lattice site. As chosen here for a square lattice, each operator has two sets of incoming modes from operators on neighboring sites and two outgoing sets of modes to operators of the remaining nearest neighbors. One outgoing set of modes corresponds to the physical site Hilbert space. The Hermitian conjugate of the circuit $\hat A_N\circ\cdots\circ\hat A_1$ is $\hat A_1^\dag\circ\cdots\circ\hat A_N^\dag$. All contraction arcs and the operator order (gray line below/above the circuit) are reversed. This side effect can be reverted (without changing the value of the FOC) by applying Eq.~\eqref{eq:operatorOrderSwap} and the rule derived in Sec.~\ref{sec:reversing} with only a marginal computational overhead.
(b) To evaluate a local expectation value, the FOCs for bra, local observable, and ket have to be composed. The operator order can again be changed for later convenience -- in this case no additional sign factors occur, as all swapped operators have no common contraction arcs. For the definition of the objects on the right hand side, see also Fig.~\ref{fig:fPEPS_b}a.}
\label{fig:fPEPS_a}
\end{figure*}
\begin{figure*}[p]
\centering
\includegraphics[width=\textwidth]{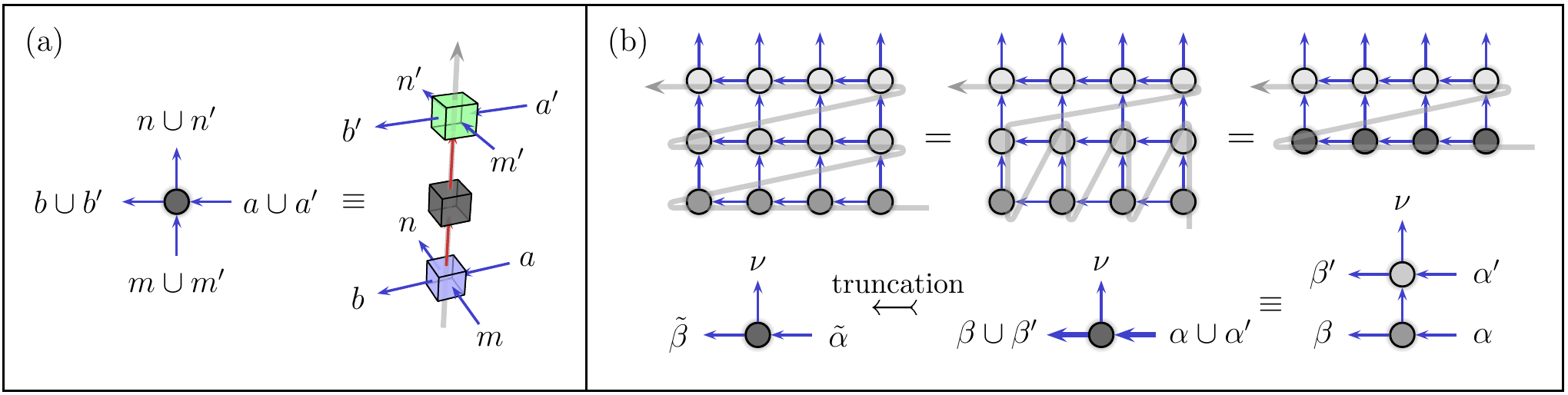}
\caption{(Color online) 
(a) Definition of the objects on the right hand side of Fig.~\ref{fig:fPEPS_a}b -- here, in particular, for the site where the local observable acts nontrivially. (b) The FOC for the evaluation of a local observable is contracted by considering the first row of the FOC as a fermionic state $\ket{\chi_1}$ and applying the other rows as operators to it $\ket{\chi_{y}}=\hat T_y \ket{\chi_{y-1}}$. Doing this in an exact manner, the number of degrees of freedom per site for the states $\ket{\chi_{y}}$ would in general increase exponentially with $y$. One can decrease them during the algorithm for the case of a finite  (infinite, translationally invariant) lattice by applying the DMRG (iTEBD) algorithm.}
\label{fig:fPEPS_b}
\end{figure*}
The FOC framework incorporates a fermionic version of the class of qudit states called \emph{tensor product ans\"atze} \cite{Niggemann1997-104,Nishino2000-575,Martin-Delgado2001-64} or  \emph{projected entangled pair states} (PEPS) \cite{Verstraete2004-7}. In Ref.\ \cite{Kraus2009_04} it was suggested to obtain fermionic PEPS by applying fermionic parity-symmetric (projection) operators to a tensor product of maximally entangled pair states. The detour over maximally entangled states is not necessary (but also not harmful); as depicted on the left hand side of Fig.~\ref{fig:fPEPS_a}a, a fermionic PEPS on a square lattice, equivalently, can be defined by assigning to each lattice site $(x,y)$ (away from the boundaries) a parity-symmetric fermionic operator $\hat A:\F_{a\cup m}\to\F_{b\cup n\cup s}$ where $a$ and $m$ are sets of incoming modes from operators on neighboring sites $(x+1,y)$ and $(x,y+1)$, and $b$ and $n$ are outgoing modes to operators on sites $(x-1,y)$ and $(x,y-1)$. The set of modes $s$ composes the local physical Hilbert space of site $(x,y)$. In the FOC framework, the generalization to more complicated or higher-dimensional lattices is straightforward. The choice of the direction of the contraction arcs is (an arbitrary) part of the definition of the state and can also be changed later as described in Sec.~\ref{sec:reversing}.
To complete the definition of the fermionic PEPS one needs to specify an (initial) operator order. An example is given on the left hand side of Fig.~\ref{fig:fPEPS_a}a, where the gray line below the lattice indicates the lexicographic order with respect to lattice coordinates $(-x,y)$.

In Ref.\ \cite{Kraus2009_04} it was described how the FOC of a fermionic PEPS can be mapped to a QUOC by choosing a fixed ordering of all modes. This was achieved with one additional bond per horizontal contraction arc (i.e., a factor of four in the number of degrees of freedom per site) and a correspondingly reduced computational efficiency (a factor of several powers of four) for the evaluation of expectation values, calculation of ground states etc. 

The approach presented here is an alternative one, emphasizing that the mapping to a QUOC (with a fixed mode order) is not necessary. All manipulations and contractions on fermionic PEPS can be done according to the rules described in Secs.~\ref{sec:contractions}, \ref{sec:operatorOrder}, and \ref{sec:furtherOp}. In that case, compared to the same operations on a corresponding qudit PEPS (replacing the fermionic operators with qudit operators of identical dimensions), only marginal computational overheads arise.

Fig.~\ref{fig:fPEPS_a} shows graphically how the FOC for the evaluation of a local expectation value $\bra{\psi}\hat O\ket{\psi}$ can be constructed. 
For the bra vector (dual vector) $\bra{\psi}$, operator order and contraction lines reverse as a side effect of taking the Hermitian conjugate; Sec.~\ref{sec:herm}. For later convenience this is reverted by applying Eq.~\eqref{eq:operatorOrderSwap} and the rule derived in Sec.~\ref{sec:reversing}. After composing bra, observable, and ket, the operator order can again be changed conveniently, this time without any sign factors occurring, as all swapped operators share no common contraction arcs; Fig.~\ref{fig:fPEPS_a}b. As in the qudit case \cite{Verstraete2004-7}, the contraction of the resulting circuit can be executed row by row, i.e., by treating the lowest row as a one-dimensional fermionic state $\ket{\chi_1}$ to which the operators of the following row $\hat T_y$ (\emph{row transfer matrix}) are applied; $\ket{\chi_{y}}=\hat T_y \ket{\chi_{y-1}}$. No additional sign factors occur due to operator reorderings (Fig.~\ref{fig:fPEPS_b}), but only due to mode reorderings (Sec.~\ref{sec:modeReordering}) before contractions (marginal overhead). An essential aspect of PEPS algorithms is that contractions, e.g., for the evaluation of expectation values, cannot be executed exactly, as the stepwise application of the row transfer matrices would in general lead to an exponential growth in the number of modes per site for $\ket{\chi_y}$. As suggested in Ref.\ \cite{Verstraete2004-7}, this can be circumvented by applying a variant of the density-matrix renormalization-group (DMRG) algorithm \cite{White1992-11,Schollwoeck2005} to each state $\ket{\chi_y}$, before executing the contractions to the next row. The only purpose of the DMRG procedure is here to reduce the number of degrees of freedom in each step to a manageable number, and hence, do contractions in an approximative fashion. The essential operation is to do Schmidt decompositions of $\ket{\chi_y}$. This can be done for FOCs as described in Sec.~\ref{sec:svd}.

The FOC framework also allows to simulate infinite fermionic PEPS. To this purpose, the fermionic PEPS is to be defined by repetition of an elementary cell FOC; cmp.\ to Ref.\ \cite{Jordan2008-101} for the qudit case. The algorithm does not deviate substantially from the finite-size case. The biggest difference being that, for the reduction of degrees of freedom in states $\ket{\chi_y}$, one has to use a translationally invariant formulation of the DMRG algorithm, basically the iTEBD algorithm as described in Ref.\ \cite{Orus2008-78}, again based on the ability to do singular value decompositions (Sec.~\ref{sec:svd}).
With this, one has a translation of the algorithms for the calculation of approximative ground state or time-evolved qudit (i)PEPS \cite{Verstraete2004-7,Jordan2008-101} to the fermionic case without reduction of the computational efficiency, as those algorithms are based on the ability to contract operator circuits just as in our example.

\section{Discussion} \label{sec:discussion}
In Ref.\ \cite{Pineda2009_05} it was shown that contractions of fermionic unitary circuits with a causal cone (for instance the evaluation of local observables w.r.t.\ a MERA) can be done without occurrence of any Jordan-Wigner strings outside the causal cone. Here, this result was extended in proving that arbitrary parity-symmetric fermionic operator circuits can actually be contracted with the \emph{same} computational effort and memory requirements as a corresponding QUOC. This remarkable result follows from the fact that a given contraction sequence for a QUOC can be implemented for a corresponding FOC with essentially the same number of computational operations. We have presented the required contraction primitives and discussed the marginal computational overheads.

This allows to translate algorithms on QUOCs to corresponding algorithms on FOCs.
For example in the algorithm for scale-invariant MERA as studied in Refs.\ \cite{Giovannetti2009-79,Montangero2008-10,Pfeifer2009-79}, the super operator simply becomes a fermionic super operator. Its iterative application to an observable yields the expectation value of the observable in the thermodynamic limit.

For the special example of the FOC being a MERA, in Ref.\ \cite{Corboz2009_04}, first numerical results where presented (postponing a description of the algorithm for a later publication). A scheme for fermionic PEPS was suggested in Ref.\ \cite{Kraus2009_04}. The suggested mapping to a QUOC used there seems numerically less efficient than the contraction scheme presented here. Instead of encoding the fermionic sign factors by increasing tensor dimensions, they can be taken account of during contractions, specifically in preparative mode reorderings, and operator order swaps. The resulting marginal overhead appears smaller.

It will be interesting to see to what extent variational ans\"atze like fermionic variants of PEPS or MERA, both satisfying entropic area laws \cite{Amico2008-80,Eisert2008,Latorre2009}, will be able to appropriately 
grasp the correlations present in critical fermionic strongly correlated models, models that are 
known to violate such area laws logarithmically \cite{Wolf2005,Gioev2005,Barthel2006-74,Li2006,Cramer2006}.
First numerical results \cite{Corboz2009_04,Kraus2009_04,Pineda2009_05} seem promising.
It is the hope that the framework discussed in this work will help in constructing fermionic variants of variational approaches to simulate strongly correlated fermions in higher dimensions.

\acknowledgments
We thank V.\ Giovannetti, M.\ Rizzi, U.\ Schollw{\"o}ck, and S.-Y.~Jang for discussions. This work has been supported by the EU (QAP, MINOS, QAP), and the EURYI.\\

\appendix

\section{Alternative derivation of the rule for partial multiplications} \label{sec:multiplyAB_alternative}
For the same preconditions as in Sec.~\ref{sec:multiplyAB}, we want to derive the partial multiplication rule -- this time by extending the supports of operators $\hat A$ and $\hat B$ prior to the contraction.

Let $\hat D:\F_m\to\F_n$, and let $\mf{m}$, $\mf{n}$, $\mf{r}$ be orderings of the modes in $m$, $n$, and $r$. Extending the incoming and outgoing supports of $\hat D$ by modes $r$ (on which the resulting operator is supposed to act trivially), we arrive at $\hat D':\F_{m\cup r}\to\F_{n\cup r}$ with
\begin{equation}
	\hat D'=\Ket{\vec{n}\vec{r}}{\mf{n}\oplus\mf{r}}
 \Bra{\vec{n}}{\mf{n}} \hat D \Ket{\vec{m}}{\mf{m}} 
 \Bra{\vec{m}\vec{r}}{\mf{m}\oplus\mf{r}}.
\label{eq:supportExtension}
\end{equation}
This is confirmed by $\Tr_r\hat D'=\hat D$, according to \eqref{eq:partialTrace}, and $\hat f_x \hat D'_\pm=\pm\hat D'_\pm \hat f_x$ for all $x\in r$.

Using the rule \eqref{eq:supportExtension} to extend the supports of the operators
$\hat A:\F_{m}\to\F_{n\cup p}$ and $\hat B:\F_{n\cup q}\to\F_{k}$ by modes $q$ and $p$, respectively, defines operators $\hat A':\F_{m\cup q}\to\F_{n\cup p\cup q}$ and $\hat B':\F_{n\cup p\cup q}\to\F_{k\cup p}$.
The partial multiplication $\hat C=\hat B\cdot_n\hat A$ amounts now simply to the usual operator product $\hat C=\hat B'\cdot \hat A'$. 
Assuming we have the two operators in representations $A=J_{\mf{a},\mf{m}}(\hat A)$ and $B=J_{\mf{k},\mf{b}}(\hat B)$ with $\mf{a}=\mf{n}\oplus\mf{p}$ and $\mf{b}=\mf{n}\oplus\mf{q}$ ($\mf{p}$ and $\mf{q}$ are orderings for the modes in $p$, and $q$),
\begin{eqnarray*}
	\hat A'&=&\Ket{\vec{n}\vec{p}\vec{q}}{\mf{a}\oplus\mf{q}}
 \Bra{\vec{n}\vec{p}}{\mf{a}} \hat A \Ket{\vec{m}}{\mf{m}} 
 \Bra{\vec{m}\vec{q}}{\mf{m}\oplus\mf{q}},
\end{eqnarray*}
and
\begin{eqnarray*}
	\hat B'&=&\Ket{\vec{k}\vec{p}}{\mf{k}\oplus\mf{p}}
 \Bra{\vec{k}}{\mf{k}} \hat B \Ket{\vec{n}\vec{q}}{\mf{b}} 
 \Bra{\vec{n}\vec{q}\vec{p}}{\mf{b}\oplus\mf{p}}\\
&=& (-1)^{\bar{p}\bar{q}}\Ket{\vec{k}\vec{p}}{\mf{k}\oplus\mf{p}}
 (\vec{k}| B |\vec{n}\vec{q}) \Bra{\vec{n}\vec{p}\vec{q}}{\mf{n}\oplus\mf{p}\oplus\mf{q}}.
\end{eqnarray*}

The result $\hat C:\F_{m\cup q}\to\F_{k\cup p}$ of the multiplication with orderings $\mf{c}_1=\mf{k}\oplus\mf{p}$, $\mf{c}_2=\mf{m}\oplus\mf{q}$ is then
\begin{eqnarray*}
	\hat C&=&\hat B'\cdot\hat A' \\
 &=& (-1)^{\bar{p}\bar{q}}
 \Ket{\vec{k}\vec{p}}{\mf{k}\oplus\mf{p}}
 (\vec{k}| B |\vec{n}\vec{q}) 
 (\vec{n}\vec{p}| A |\vec{m}) \Bra{\vec{m}\vec{q}}{\mf{m}\oplus\mf{q}} \\
 &=& \Ket{\vec{k}\vec{p}}{\mf{c}_1}(\vec{k}\vec{p}|C|\vec{m}\vec{q})\Bra{\vec{m}\vec{q}}{\mf{c}_2},
\end{eqnarray*}
coinciding with Eq.~\eqref{eq:partialMultiply}.

\bibliographystyle{prsty} 

\end{document}